\input harvmac
\input amssym

%\draftmode

\def\bl{\bar{\lambda}}
\def\bz{\Bbb Z}
\def\wt{\widetilde}

%%%%%%%%%%%%% For the Figures %%%%%%%%%%%%%%%%%%%%%
\input epsf.tex
\def\figin{\epsfcheck\figin}\def\figins{\epsfcheck\figins}
\def\epsfcheck{\ifx\epsfbox\UnDeFiSIeD
\message{(NO epsf.tex, FIGURES WILL BE IGNORED)}
\gdef\figin##1{\vskip2in}\gdef\figins##1{\hskip.5in}% blank space instead
\else\message{(FIGURES WILL BE INCLUDED)}%
\gdef\figin##1{##1}\gdef\figins##1{##1}\fi}
\def\DefWarn#1{}
\def\figinsert{\goodbreak\midinsert}
\def\ifig#1#2#3{\DefWarn#1\xdef#1{fig.~\the\figno}
\writedef{#1\leftbracket fig.\noexpand~\the\figno}%
\figinsert\figin{\centerline{#3}}\medskip\centerline{\vbox{\baselineskip12pt
\advance\hsize by -1truein\noindent\footnotefont{\bf
Fig.~\the\figno:} #2}}
\bigskip\endinsert\global\advance\figno by1}
%%%%%%%%%%%%%%%%%%%%%%%%%%%%%%%%%%%%%%%%%%%%%%%%%%%%

%%%%%%%%%%%%%%%%%%% References %%%%%%%%%%%%%%%%%%%%%%%%%%%%%%
%\SeibergEB
\lref\SeibergEB{
  N.~Seiberg,
  ``Notes On Quantum Liouville Theory And Quantum Gravity,''
  Prog.\ Theor.\ Phys.\ Suppl.\  {\bf 102}, 319 (1990).
  %%CITATION = PTPSA,102,319;%%
}

%\GiveonZZ
\lref\GiveonZZ{
 A.~Giveon and A.~Sever,
 ``Strings in a 2-d extremal black hole,''
 JHEP {\bf 0502}, 065 (2005)
 [arXiv:hep-th/0412294].
 %%CITATION = HEP-TH 0412294;%%
}

%\PolchinskiRR
\lref\PolchinskiRR{
   J.~Polchinski,
   ``String theory. Vol. 2: Superstring theory and beyond,''
   Cambridge University Press (1998).
%\href{http://www.slac.stanford.edu/spires/find/hep/www?
%irn=4634802}{SPIRES entry}
}

%\DouglasUP
\lref\DouglasUP{
   M.~R.~Douglas, I.~R.~Klebanov, D.~Kutasov, J.~Maldacena, E.~Martinec
and N.~Seiberg,
   ``A new hat for the c = 1 matrix model,''
   [arXiv:hep-th/0307195].
   %%CITATION = HEP-TH 0307195;%%
}

%\KlebanovQA
\lref\KlebanovQA{
   I.~R.~Klebanov,
   ``String theory in two-dimensions,''
   arXiv:hep-th/9108019.
   %%CITATION = HEP-TH 9108019;%%
}

%\GinspargIS
\lref\GinspargIS{
   P.~H.~Ginsparg and G.~W.~Moore,
   ``Lectures on 2-D gravity and 2-D string theory,''
   arXiv:hep-th/9304011.
   %%CITATION = HEP-TH 9304011;%%
}

%\PolchinskiMB
\lref\PolchinskiMB{
   J.~Polchinski,
   ``What is string theory?,''
   arXiv:hep-th/9411028.
   %%CITATION = HEP-TH 9411028;%%
}

%\MaldacenaHE
\lref\MaldacenaHE{
   J.~Maldacena and N.~Seiberg,
   ``Flux-vacua in two dimensional string theory,''
   JHEP {\bf 0509}, 077 (2005)
   [arXiv:hep-th/0506141].
   %%CITATION = HEP-TH 0506141;%%
}

%\KarczmarekPV
\lref\KarczmarekPV{
  J.~L.~Karczmarek and A.~Strominger,
  ``Matrix cosmology,''
  JHEP {\bf 0404}, 055 (2004)
  [arXiv:hep-th/0309138].
  %%CITATION = HEP-TH 0309138;%%
}

%\KarczmarekPH
\lref\KarczmarekPH{
  J.~L.~Karczmarek and A.~Strominger,
  ``Closed string tachyon condensation at c = 1,''
  JHEP {\bf 0405}, 062 (2004)
  [arXiv:hep-th/0403169].
  %%CITATION = HEP-TH 0403169;%%
}

%\DasHW
\lref\DasHW{
  S.~R.~Das, J.~L.~Davis, F.~Larsen and P.~Mukhopadhyay,
  ``Particle production in matrix cosmology,''
  Phys.\ Rev.\ D {\bf 70}, 044017 (2004)
  [arXiv:hep-th/0403275].
  %%CITATION = HEP-TH 0403275;%%
}

%\DasAQ
\lref\DasAQ{
  S.~R.~Das and J.~L.~Karczmarek,
  ``Spacelike boundaries from the c = 1 matrix model,''
  Phys.\ Rev.\ D {\bf 71}, 086006 (2005)
  [arXiv:hep-th/0412093].
  %%CITATION = HEP-TH 0412093;%%
}

%\DineVU
\lref\DineVU{
  M.~Dine, P.~Y.~Huet and N.~Seiberg,
  ``Large And Small Radius In String Theory,''
  Nucl.\ Phys.\ B {\bf 322}, 301 (1989).
  %%CITATION = NUPHA,B322,301;%%
}

%\KutasovUA
\lref\KutasovUA{
   D.~Kutasov and N.~Seiberg,
   ``Noncritical Superstrings,''
   Phys.\ Lett.\ B {\bf 251}, 67 (1990).
   %%CITATION = PHLTA,B251,67;%%
}

%%%%%%%%%%%%%%%%% Discrete States %%%%%%%%%%%%%%%%%%%%%%%%%%%%%%%%%%
%\ItohIY
\lref\ItohIY{ K.~Itoh and N.~Ohta, ``BRST cohomology and physical
states in 2-D
supergravity coupled to $c \le 1$ matter,'' Nucl.\ Phys.\ B {\bf 377},
113
(1992) [arXiv:hep-th/9110013].
%%CITATION = HEP-TH 9110013;%%
}

%\BouwknegtVA
\lref\BouwknegtVA{ P.~Bouwknegt, J.~G.~McCarthy and K.~Pilch, `BRST
analysis of
physical states for 2-D (super)gravity coupled to (super)conformal
matter,''
arXiv:hep-th/9110031.
%%CITATION = HEP-TH 9110031;%%
}

%\BouwknegtAM
\lref\BouwknegtAM{ P.~Bouwknegt, J.~G.~McCarthy and K.~Pilch, ``Ground
ring for
the 2-D NSR string,'' Nucl.\ Phys.\ B {\bf 377}, 541 (1992)
[arXiv:hep-th/9112036].
%%CITATION = HEP-TH 9112036;%%
}

%\ItohIX
\lref\ItohIX{ K.~Itoh and N.~Ohta, ``Spectrum of two-dimensional
(super)gravity,'' Prog.\ Theor.\ Phys.\ Suppl.\  {\bf 110}, 97 (1992)
[arXiv:hep-th/9201034].
%%CITATION = HEP-TH 9201034;%%
}

%%%%%%%%%%%%%%%%%%%%%%%%%%%%%%%%%%%%%%%%%%%%%%%%%%%%%%%%%%%%%%%%%%%%%%

%\McGuiganQP
\lref\McGuiganQP{
   M.~D.~McGuigan, C.~R.~Nappi and S.~A.~Yost,
   ``Charged black holes in two-dimensional string theory,''
   Nucl.\ Phys.\ B {\bf 375}, 421 (1992)
   [arXiv:hep-th/9111038].
   %%CITATION = HEP-TH 9111038;%%
}

%\TakayanagiSM
\lref\TakayanagiSM{
   T.~Takayanagi and N.~Toumbas,
``A matrix model dual of type 0B string theory in two dimensions,''
   JHEP {\bf 0307}, 064 (2003)
   [arXiv:hep-th/0307083].
   %%CITATION = HEP-TH 0307083;%%
}

%\GukovYP
\lref\GukovYP{
   S.~Gukov, T.~Takayanagi and N.~Toumbas,
``Flux backgrounds in 2D string theory,''
   JHEP {\bf 0403}, 017 (2004)
   [arXiv:hep-th/0312208].
   %%CITATION = HEP-TH 0312208;%%
}

%\TakayanagiGE
\lref\TakayanagiGE{
T.~Takayanagi,
``Comments on 2D type IIA string and matrix model,''
JHEP {\bf 0411}, 030 (2004)
[arXiv:hep-th/0408086].
%%CITATION = HEP-TH 0408086;%%
}

%\SeibergBX
\lref\SeibergBX{
   N.~Seiberg,
   ``Observations on the moduli space of two dimensional string theory,''
   [arXiv:hep-th/0502156].
   %%CITATION = HEP-TH 0502156;%%
}

%\DavisQE
\lref\DavisQE{
  J.~L.~Davis, F.~Larsen and N.~Seiberg,
  ``Heterotic strings in two dimensions and new stringy phase transitions,''
  JHEP {\bf 0508}, 035 (2005)
  [arXiv:hep-th/0505081].
  %%CITATION = HEP-TH 0505081;%%
}

%\SeibergNK
\lref\SeibergNK{
  N.~Seiberg,
  ``Long Strings, Anomaly Cancellation, Phase Transitions, T-duality and
  Locality in the 2d Heterotic String,''
  [arXiv:hep-th/0511220].
  %%CITATION = HE-TH 0511220;%%
}

%\BergmanYP
\lref\BergmanYP{
   O.~Bergman and S.~Hirano,
   ``The cap in the hat: Unoriented 2D strings and matrix(-vector)
models,''
   JHEP {\bf 0401}, 043 (2004)
   [arXiv:hep-th/0311068].
   %%CITATION = HEP-TH 0311068;%%
}

%\GomisVI
\lref\GomisVI{
   J.~Gomis and A.~Kapustin,
   ``Two-dimensional unoriented strings and matrix models,''
   JHEP {\bf 0406}, 002 (2004)
   [arXiv:hep-th/0310195].
   %%CITATION = HEP-TH 0310195;%%
}

%\KlebanovKM
\lref\KlebanovKM{
   I.~R.~Klebanov, J.~Maldacena and N.~Seiberg,
   ``D-brane decay in two-dimensional string theory,''
   JHEP {\bf 0307}, 045 (2003)
   [arXiv:hep-th/0305159].
   %%CITATION = HEP-TH 0305159;%%
}

\lref\McGreevyKB{
   J.~McGreevy and H.~Verlinde,
   ``Strings from tachyons: The c = 1 matrix reloaded,''
   JHEP {\bf 0312}, 054 (2003)
   [arXiv:hep-th/0304224].
   %%CITATION = HEP-TH 0304224;%%
}

%\GomisCE
\lref\GomisCE{
  J.~Gomis,
  ``Anomaly cancellation in noncritical string theory,''
  JHEP {\bf 0510}, 095 (2005)
  [arXiv:hep-th/0508132].
  %%CITATION = HEP-TH 0508132;%%
}

%\AlvarezSJ
\lref\AlvarezSJ{
   E.~Alvarez and M.~A.~R.~Osorio,
   ``Superstrings At Finite Temperature,''
   Phys.\ Rev.\ D {\bf 36}, 1175 (1987).
   %%CITATION = PHRVA,D36,1175;%%
}

%\BrienPN
\lref\BrienPN{
   K.~H.~O'Brien and C.~I.~Tan,
   ``Modular Invariance Of Thermopartition Function And Global Phase
Structure
   Of Heterotic String,''
   Phys.\ Rev.\ D {\bf 36}, 1184 (1987).
   %%CITATION = PHRVA,D36,1184;%%
}

%\PolchinskiZF
\lref\PolchinskiZF{
   J.~Polchinski,
   ``Evaluation Of The One Loop String Path Integral,''
   Commun.\ Math.\ Phys.\  {\bf 104}, 37 (1986).
   %%CITATION = CMPHA,104,37;%%
}

%%%%%%%%%%%%%%%%%% Title Page %%%%%%%%%%%%%%%%%%%%%%%%%%%%%

\Title{
}
{\vbox{\centerline{The Moduli Space and Phase Structure}
\medskip
\centerline{of Heterotic Strings in Two Dimensions}}}
\medskip
\centerline{\it
Joshua L. Davis
}
\bigskip
\centerline{Michigan Center for Theoretical Physics,}
\smallskip
\centerline{University of Michigan, Ann Arbor, MI 48109}
\smallskip

\vglue .3cm
%\vskip 2cm
\bigskip\bigskip\bigskip
\centerline{\bf Abstract}

\noindent
We explore the moduli space of heterotic strings in two dimensions. In doing so, we introduce new lines of compactified theories with $Spin(24)$ gauge symmetry and discuss compactifications with Wilson lines. The phase structure of $d=2$ heterotic string theory is examined by classifying the hypersurfaces in moduli space which support massless quanta or discrete states. Finally, we compute the torus amplitude over much of the moduli space.

\Date{}

%%%%%%%%%%%%%%%%%%%% Bulk %%%%%%%%%%%%%%%%%%%%%%%%%%%%%%%%%%%

\newsec{Introduction}

The past few years have seen a resurgence of interest in two-dimensional string theory. Just as in the first period of interest in $d=2$ strings (for a review see \refs{\KlebanovQA\GinspargIS-\PolchinskiMB}), the latest studies have used these theories as toy models of phenomena in higher dimensions. Topics of recent interest include tachyon condensation and D-brane decay \refs{\McGreevyKB,\KlebanovKM}, flux vacua \refs{\GukovYP,\MaldacenaHE} and time-dependent backgrounds \refs{\KarczmarekPV\KarczmarekPH\DasHW-\DasAQ}.

The current research program on strings in two dimensions differs from the work of the early 1990's in an important aspect. The vast majority of the work of that earlier period focused on the $c=1$ bosonic string theory. In contrast, there are many $d=2$ models currently being discussed in the literature \refs{\TakayanagiSM\DouglasUP\GomisVI\BergmanYP\TakayanagiGE-\SeibergBX}. This ``landscape'' of two-dimensional theories should be further explored. It would be of interest to develop the entire web of dualities and interrelations of these theories; some work in this direction includes \SeibergBX.

A less explored region of this landscape consists of theories with ${\cal N} = (1,0)$ worldsheet supersymmetry, {\it i.e.} heterotic strings. These models are of particular interest because they manifest a range of phenomena which is intermediate in richness and scope; they are more complex and interesting than their two-dimensional cousins but still more tractable than higher-dimensional string theories. For example, unlike other $d=2$ string theories, these theories are known to exhibit Hagedorn-like phase transitions and, unlike Hagedorn behavior in higher dimensions, one can retain computational control on both sides of the transition \DavisQE. Additionally, the heterotic theories seem to be related to two-dimensional black holes \refs{\McGuiganQP,\GiveonZZ} which has been difficult to study with other $d=2$ strings.

%%%%%%%%%%%%%%%%%%%%%%%%%%%%%%%%%%%%%%%%%%%%%%% New Material %%%%%%%%%%%%%%%%%%%%%%%%%%%%%%%%%%%%%%%%%%%%%%%%%%%%%%%%%
The purpose of the present paper is to fully explore the moduli space of heterotic strings discovered in \DavisQE. In particular, we wish to compute the one-loop free energy (torus amplitude) over this entire space. However, we find this endeavor to be obstructed at small compactification radius by the existence of a rich network of surfaces of first-order phase transitions. The locations of such surfaces are governed by the appearance of massless modes localized on moduli space. We classify the spectrum of these modes as a function of the moduli space coordinates in order to map out the phase structure of the theory. Additionally we catalogue the spectrum of discrete states on moduli space, and in so doing, chart the regions of enhanced gauge symmetry.
%%%%%%%%%%%%%%%%%%%%%%%%%%%%%%%%%%%%%%%%%%%%%%%%%%%%%%%%%%%%%%%%%%%%%%%%%%%%%%%%%%%%%%%%%%%%%%%%%%%%%%%%%%%%%%%%%%%%%%

\subsec{Summary}
%%%%%%%%%%%%%%%%%%%%%%%%%%%%%%%%%%%%%%%%%%%%Added Sentence %%%%%%%%%%%%%%%%%%%%%%%%%%%%%%%%%%%%%%%%%%%%%%%%%%%%%%
Due to the technical nature of the analysis, let us briefly summarize the main features of the $d=2$ heterotic moduli space found in \DavisQE\ and this work.
%%%%%%%%%%%%%%%%%%%%%%%%%%%%%%%%%%%%%%%%%%%%%%%%%%%%%%%%%%%%%%%%%%%%%%%%%%%%%%%%%%%%%%%%%%%%%%%%%%
There are three distinct non-compact limits. The ``heterotic orthogonal'' (HO) theory is a $Spin(24)$ gauge theory with a single massless scalar in the fundamental of the gauge group. The ``heterotic exceptional'' (HE) theory has a $Spin(8) \times E_8$ gauge group. The propagating modes are $E_8$ singlets and charged under the $Spin(8)$ as follows: a scalar in the ${\bf 8}_v$ and two chiral fermions in the ${\bf 8}_s$ and ${\bf 8}_c$ irreps. The HO and HE theories were introduced in \McGuiganQP\ and explored in more detail in \refs{\GiveonZZ,\DavisQE}. In addition, another theory which we shall dub ``twisted HO'' (THO) was introduced in \DavisQE. It was mentioned only briefly there for it appears to be anomalous; it is a $Spin(24)$ gauge theory with spacetime spectrum consisting of a single chiral fermion in the fundamental representation. However, this theory is entirely consistent from the string worldsheet point of view; for a treatment of the anomaly see \SeibergNK. In this work we develop the THO theory, including the discussion of twisted compactifications, T-duality and enhanced gauge symmetry. We then extend the analysis of phase transitions to the entire $d=2$ heterotic moduli space by discussing compactifications with arbitrary Wilson lines.

%%%%%%%%%%%%%%%%%%%%%%%%%%%%%%%%%%%%%%%%% Added material %%%%%%%%%%%%%%%%%%%%%%%%%%%%%%%%%%%%%%%%%
As mentioned above the non-compact theories have gauge group $Spin(24)$ or $Spin(8) \times E_8$, both of rank 12. In all theories, the Liouville direction must be non-compact. Thus the moduli space is 13-dimensional, parameterized by the radius $R$ of compact Euclidean time as well as the twelve component Wilson line, $R\vec{A}$, which wraps this direction (all three non-compact theories lie on this same contiguous space as shown in \DavisQE). Due to the diagonal nature of the GSO projection for these theories (which correlates the Lorentz and gauge representations), certain Wilson line backgrounds actually implement a thermal twist about the Euclidean time direction and so describe the corresponding non-compact theories in the canonical ensemble. For generic Wilson lines, more formal but still well-defined twisted compactifications arise.

Typically the twisted compactifications, including the thermal theories, have first-order transitions at specific values of $R$ due to the appearance of localized massless modes. We classify the surfaces in moduli space supporting such modes in equation (5.4), effectively providing the transition temperature as a function of Wilson line. We find regions with enhanced gauge symmetry by examining the surfaces (5.7) supporting discrete states.

We compute the torus amplitude for generic Wilson line and large radius (low temperature). It is always well-defined and in agreement with field theory predictions at large radius, however the phase transitions have the effect of preventing calculability at sufficiently small radius (large temperature) except in cases where T-duality can be exploited. At low temperatures the thermodynamics is standard, but at high temperatures the worldsheet analysis of this article is insufficient to capture the entire physics (see \SeibergNK\ for discussion of the high temperature phase of twists with unbroken gauge symmetry). The torus amplitudes for large radii are given for the HO, HE and THO theories with Wilson lines by equations (6.17), (6.18) and (6.19) respectively. These expressions are valid for the regions of moduli space with $R$ above all of the transitions. We discuss the formal obstructions to evaluating the torus amplitude below the transition in section 6.3.
%%%%%%%%%%%%%%%%%%%%%%%%%%%%%%%%%%%%%%%%%%%%%%%%%%%%%%%%%%%%%%%%%%%%%%%%%%%%%%%%%%%%%%%%%%%%%%%%%%%

The layout of this paper is as follows. In section 2, we introduce our notation and discuss the discrete symmetries and general features of ${\cal N} = (1,0)$ worldsheet conformal field theory with two-dimensional target space.  Section 3 contains a short review of the three theories with non-compact target space: HO, HE and THO. This is followed by the development of the most symmetric compactifications of THO in section 4, {\it i.e.} those which leave the $Spin(24)$ gauge group unbroken. The spectra, T-duality properties and points of enhanced symmetry are all described. In section 5 we introduce compactifications with Wilson lines and map out the phase structure by classifying the regions of moduli space which support massless quanta or discrete states. Finally, in section 6, we calculate the torus amplitude throughout much of moduli space and discuss the obstacle in continuing this computation to the entire space.
Throughout the paper we will use conventions in which $\alpha^\prime =2$.

\newsec{Worldsheet Theory and Discrete Symmetries}

All of the theories to be considered in this work have the same ${\cal N} = (1,0)$ worldsheet field content; the differences arise from the choice of GSO projection. The right-moving CFT is that of the $\hat{c}=1$ noncritical superstring: a Liouville field $\phi$, (Euclidean) time $x$, and their fermionic superpartners $\psi_\phi$ and $\psi_x$. The slope of the Liouville field is such that it contributes $c_{\phi}=13$ to the central charge and the string coupling varies as $e^\phi$. The left-moving CFT again includes the Liouville field and Euclidean time, as well as a $c=12$ current algebra which can be represented by 24 free fermions $\bar{\lambda}^I$ with $I=1 \ldots 24$. It is useful to bosonize the worldsheet fermions as follows:
\eqn\bosonize{\eqalign{
e^{\pm i H} & = { \psi_x \pm i \psi_\phi  \over \sqrt{2} }\cr
e^{\pm i (\vec{H}_L)_k} & = { \bar{\lambda}^{2k-1} \pm i \bar{\lambda}^{2k} \over \sqrt{2} } ~~ , \quad k=1 \ldots 12
}}
The operators in the current algebra can then be classified by the lattice of $\vec{H}_L$ momentum. This will be discussed further shortly.

To develop our theories let us first introduce a set of discrete symmetry generators. In the right moving sector define $F_R$, the right-moving spacetime fermion number, and $f_R$, the right-moving worldsheet fermion number, as in \SeibergBX,
\eqn\symmR{\eqalign{
& (-)^{F_R}: ~~~\varphi\to \varphi + 2\pi i~~\cr
& (-)^{f_R}:~~~\varphi\to \varphi +\pi i~,~~H \to H+\pi
}}
In the above, $\varphi$ refers to the bosonized field in the $\beta\gamma$ superconformal ghost system (see {\it e.g.} \PolchinskiRR). It is clear that $F_R$ and $f_R$ are defined only {\it mod 2}.

On the left-moving side $F_L$ and $f_L$, the left-moving spacetime and worldsheet fermion numbers, are defined using the group properties of the $c=12$ current algebra. The $\bar{\lambda}$ can be arranged into either a $Spin(24)$ or $Spin(8) \times E_8$ affine Lie algebra \refs{\McGuiganQP,\DavisQE}. In the $Spin(8) \times E_8$ theories, we will only consider states with an even number of $E_8$ fermions excited. With this condition $F_L$ and $f_L$ are defined by their action on $Spin(2n)$ conjugacy classes. The center of $Spin(2n)$ is $\bz_2 \times \bz_2 $ where a convenient choice of basis is given by $\left( (-)^{F_L},(-)^{f_L}\right)$. These operators act as
\eqn\symmL{\eqalign{
& (-)^{F_L}: {\cal O }_{0/V} \to {\cal O }_{0/V}, \quad {\cal O }_{C/S} \to -{\cal O }_{C/S} \cr
& (-)^{f_L}: {\cal O }_{0/S} \to {\cal O }_{0/S}, \quad {\cal O }_{V/C} \to -{\cal O }_{V/C}
}}
where ${\cal O}_r$ is some operator in the $r$ conjugacy class of $Spin(2n)$, {\it i.e.} whose bosonized form has $\vec{H}_L$ momentum in the $r$ weight lattice of $Spin(2n)$. The $0$ \& $V$, or root and vector, lattices have elements $\vec{\omega}$ such that $\omega_k \in \bz$ and the $S$ \& $C$, or spinor and anti-spinor, lattices contain  $\omega_k \in \bz +\half$. Additionally, the $0/S$ ($V/C$) lattices have elements such that $\sum \omega_k \in 2 \bz$ ($2 \bz + 1$). The definitions \symmL\ match with the usual notion that $(-1)^{F_L}$ even (odd) states are in the left-moving Neveu-Schwarz (Ramond) sector and that $f_L$ counts the sum of the momenta in the bosonized language. As with their right-moving counterparts, $F_L$ and $f_L$ are defined only {\it mod 2}.

It is worth pointing out that ``left-moving spacetime fermion number'' is a bit of a misnomer. Since the fields $\bl^I$ carry no Lorentz quantum numbers their excitations do not contribute to the spacetime spin of a state. Thus the true spacetime fermion number is counted purely by $(-1)^{F_R}$.

Finally, consider the parity and charge conjugation transformations, ${\cal P}$ and ${\cal C}$. These operators act
as
\eqn\cp{\eqalign{
&{\cal P}:~ H \to -H, \quad x \to -x, \quad \bar{x} \to -\bar{x} \cr
&{\cal C}:\quad \CO_S \leftrightarrow \CO_C
}}
These operators do not necessarily generate symmetries. For later use observe that $(-1)^{F_L+f_L}$ and $(-1)^{f_L}$ are related by ${\cal CP}$ conjugation.

\newsec{The Non-compact Theories}

There are three basic theories of heterotic strings in two non-compact dimensions: HO, HE, and twisted HO (or THO). The HE and HO theories were first discussed in \McGuiganQP\ and developed further in \DavisQE. The THO theory is mentioned briefly in \DavisQE, where it appears as an orbifold of HO. Here we will construct the THO theory without recourse to HO.

\subsec{HO}

The HO theory is the simplest of the three theories. It is defined via the diagonal GSO projection $(-1)^{F_L+F_R} = (-1)^{f_L+f_R} =1$. The resulting spacetime spectrum is given by the vertex operators,
 \eqn\hospect{\eqalign{
  &G={\cal J} \bar {\cal J}\cr
  &A^{IJ}=  {\cal J}\bar \lambda^I \bar\lambda^J\cr
  &T^I(k)=e^{-\varphi}\bar \lambda^I V_k
   }}
  where the operators
  \eqn\uone{\eqalign{
   &{\cal J}=e^{-\varphi}\psi_x\cr
  &\bar{\cal  J}=\bar \partial \bar x
}}
are $U(1)$ currents and the wave functions are
  \eqn\vldef{
  V_k = e^{i k(x+\bar x) +
  (1-|k|)(\phi+\bar \phi)}
}
This is a two-dimensional $Spin(24)$ gauge theory where the $T^I(k)$ are the only propagating particle-like degrees of freedom, a scalar in the fundamental representation. The graviton, $G$, and the gauge field, $A^{IJ}$, are examples of discrete states, and exist only at zero momentum. The Ramond sector of the $\bar{\lambda}$ current algebra does not result in any physical particles, because the $Spin(24)$ spin fields are too heavy to level match with the Ramond ground states of the right-moving CFT. One can observe from \hospect\ that both ${\cal P}$ and ${\cal C}$ are symmetries.

The contribution to the partition function from the current algebra fermions is
\eqn\hopart{ Z_{HO}^{(\lambda)} (\bar{\tau}) = \half \left[ Z_0^0
(\tau)^{12}  - Z_1^0 (\tau)^{12} -
Z^1_0(\tau)^{12} \right]^* } which is modular invariant. Note that $Z_{HO}^{(\lambda)} (\bar{\tau}) = 24$ identically
\GiveonZZ.

\subsec{HE}

This theory is constructed by first dividing the worldsheet fermions, $\bar{\lambda}^I$, into two groups. The first 8 fermions, denoted $\bar{\lambda}^i$ with $i=1 \ldots 8$, form a $Spin(8)$ current algebra and the other sixteen are arranged into an $E_8$ CFT. We will require that all states have an even number of $E_8$ fermions. Then HE has the same diagonal GSO projection as HO. The spacetime spectrum is
\eqn\hespect{\eqalign{
  &T^i(k)=e^{-\varphi}\bar \lambda^i V_k \cr
  &\Psi^\alpha=e^{-\half\varphi+i\half H}\bar S^\alpha V_k~~,\quad k\geq
0 \cr
  &\wt\Psi^{\dot\alpha}=e^{-\half\varphi-i\half H}\bar
  S^{\dot\alpha} V_k ~~,\quad k\leq 0
  }}
Additionally, there is the graviton $G={\cal J} \bar{\cal J}$ and appropriate gauge fields constructed from the current algebra. The propagating spectrum includes a scalar, $T^i(k)$, in the $Spin(8)$ fundamental representation and two chiral fermions, $\Psi^\alpha$ and $\wt\Psi^{\dot\alpha}$, which transform in the ${\bf 8}_s$ and ${\bf 8}_c$ spinor representations of $Spin(8)$, respectively. The constraint on the fermion momentum is derived from BRST invariance (specifically the Dirac equation). Neither ${\cal P}$ nor ${\cal C}$ are symmetries but the combination ${\cal CP}$ is.

The contribution to the partition function from the current algebra fermions is
\eqn\hepart{\eqalign{ Z_{HE}^{(\lambda)}(\bar{\tau}) &= \half \left[ Z_0^0
({\tau})^{4}  - Z_1^0 ({\tau})^{4} - Z_0^1
({\tau})^{4} \right]^* \cdot\left( Z_0^0 ({\tau})^8 + Z_0^1
({\tau})^8 + Z_1^0 ({\tau})^8 \right)^*\cr &= [(8-8) +
(64-64)\bar{q}+\cdots ]\cdot [1 + 248\bar{q}+\cdots]=0 }}
This modular invariant expression vanishes identically due to the Jacobi identity for theta-functions. This is familiar from the vanishing of the partition functions of ten-dimensional string theories with spacetime supersymmetry. However, in the case of the HE theory there is no spacetime supersymmetry but rather a coincidental cancellation of vacuum energies.

\subsec{THO}

This theory can be described by the GSO projection $(-)^{f_L+f_R} = (-)^{F_L + F_R + f_L} =1$. The non-compact THO theory has as its only propagating mode
\eqn\thospect{
\wt\Psi^I = e^{-\half\varphi-i\half H} \bar{\lambda}^I V_k~~,\quad
k\leq 0
}
which is a chiral fermion in the fundamental of $Spin(24)$. Additionally, this theory has the discrete states, $G$ and $A^{IJ}$, defined in \hospect. None of the spacetime transformations ${\cal C}$, ${\cal P}$ or ${\cal CP}$ generate symmetries of THO. The modular invariant partition function is
\eqn\thopart{
Z_{THO}^{(\lambda)}(\bar{\tau}) = - {1 \over 4} \left[Z_0^0 ({\tau})^{12}  - Z_1^0 ({\tau})^{12} - Z^1_0({\tau})^{12} \right]^* = -12
}
which is precisely $-\half$ that of the partition function for the HO theory. The $-\half$ can be understood as follows: the THO theory is roughly the same as HO but with a chiral fermion instead of a scalar. The sign is due to fermionic statistics and the factor of $\half$ from the constraint that $k\leq 0$.

It is immediately clear that the spectrum \thospect\ is anomalous for it consists of only a chiral fermion in the fundamental representation. However, this theory is entirely consistent from the worldsheet point of view; the spectrum is mutually local and the partition function is modular invariant. This is not unlike the case of the $d=10$ heterotic superstring theories. The solution there is the Green-Schwarz mechanism and we might expect a similar resolution here. This is indeed the case; for a full treatment of this anomaly see \SeibergNK. For the purpose of this work we will assume that the THO is a consistent theory.

As a final point, note that the THO theory can be understood as the $(-1)^{f_L}$ orbifold of the HO theory. There is no distinct ``THE'' theory since such an orbifold of the HE theory is equivalent to HE via $Spin(8)$ triality \DavisQE.

\newsec{THO Compactifications}

In addition to the non-compact theories, one can examine backgrounds with compact Euclidean time. One cannot compactify the Liouville field; the linear dilaton background is not periodic. The most symmetric compactifications of HO and HE were studied in \DavisQE; we will now do the same for THO. There are four possible compactifications which preserve the full gauge group. These are compactifications with a twist by some element of the center of $Spin(24)$. Note this is somewhat different from the HO and HE theories, which had only three such distinct compactifications. This follows from the ${\cal CP}$ invariance of those theories and the ${\cal CP}$ equivalence of $(-1)^{F_L+f_L}$ and $(-1)^{f_L}$.

\subsec{Circle theory}

One can compactify THO on a circle of radius $R$ without a twist to obtain the spectrum
\eqn\thocirc{\eqalign{
  &T_C=e^{-\varphi}\bar \CO_C(\half +nw)
V_{n,w}\cr
  &\Psi_S = e^{-\half\varphi + i\half H }\bar \CO_S(\half +nw) V_{n,w}~,
\qquad  p_R \ge 0  \cr
  &\wt \Psi_V = e^{-\half\varphi - i\half H }\bar \CO_V(\half +nw)
V_{n,w}~, \qquad p_R  \le 0\cr
  }}
  The compact wave function is given by
\eqn\wavef{
V_{n,w}=
e^{i {n\over R}(x+{\bar x}) + i {w\over 2} R(x-{\bar x}) +
(1-|p_R|)(\phi + {\bar \phi})}
}
where $p_{R} = {n\over R} + {wR\over 2}$ and $n,w \in \bz$. The notation $\CO_r (\bar{\Delta})$ refers to an operator in the $\bl$ current algebra of conformal dimension $\bar{\Delta}$ and with lattice vector in the $r$ conjugacy class of $Spin(24)$.

In the type 0 and type II string theories in $d=2$ \refs{\DouglasUP, \SeibergBX} there are only states with non-zero momentum {\it or} non-zero winding but not both. This is due to the nearly trivial BRST cohomology of the $\hat{c}=1$ string \refs{\ItohIY\BouwknegtVA-\BouwknegtAM}, which requires both sides to be in the ground state for propagating particles. Thus $p_L^2 = p_R^2$ which is satisfied only by pure winding or momentum states. In the heterotic theories under study here, it is true that the right-moving CFT has the same nearly trivial cohomology but the left-moving CFT can be excited to arbitrary level in the way one is accustomed to in higher dimensions. Thus there can be a mismatch between $p_L^2$ and $p_R^2$, and so there exist states of both winding and momentum. This leads to a large, Hagedorn-like, density of states as in the critical string.

It is clear that the above spectrum is invariant under the T-duality transformation $R \to {2 \over R}$ (with the simultaneous exchange of momentum and winding, $n \leftrightarrow w$). The left-moving $U(1)$ symmetry generated by $\bar{\cal J}= \bar{\partial} \bar{x}$ is enhanced to $SU(2)$ at $R = \sqrt{2}$ in the usual way for a compactified free boson.

The torus amplitude for this theory is
\eqn\circpart{
Z_{T^2} = -{V_L \over 2} \left(R + {2 \over R} \right)
}
where $V_L$ is the volume of the weak coupling region of the Liouville direction. This can be obtained from the more general calculation later in this paper. Note that \circpart\ respects the self-duality of this theory under $R \to { 2 \over R}$.

\subsec{ $(-1)^{F_L}$ twist}

The first non-trivial compactification of the THO theory we will study has a twist by $(-1)^{F_L}$ around the Euclidean time circle. The spectrum is
\eqn\thovect{\eqalign{
  &T_C=e^{-\varphi}\bar \CO_C(\half +(2n+1)w) V_{n+\half,2w}
  \cr
  &T_S=e^{-\varphi}\bar \CO_S(\half +n(2w+1)) V_{n,2w+1}
  \cr
  &\Psi_S = e^{-\half\varphi + i\half H }\bar \CO_S(
  \half +(2n+1)w) V_{n+\half,2w}~,\qquad ~~~~~~~p_R \geq 0  \cr
  &\Psi_C = e^{-\half\varphi + i\half H }\bar \CO_C(
  \half +n(2w+1)) V_{n,2w+1}~,\qquad ~~~~~~~p_R \geq 0  \cr
  &\wt \Psi_0 = e^{-\half\varphi - i\half H }\bar \CO_0(
  \half +(n+\half)(2w+1)) V_{n+\half,2w+1}~, \quad p_R \leq 0  \cr
  &\wt \Psi_V = e^{-\half\varphi - i\half H }
  \bar \CO_V(\half +2nw) V_{n,2w}~, \qquad \qquad ~~~~~~~~~~~~p_R \leq 0 \cr
  }}
The momentum and winding of operators in this theory are correlated with the $Spin(24)$ conjugacy classes. In the untwisted sector (consisting of even winding), operators which are even (odd) under $(-1)^{F_L}$ have integer (half-integer) momenta around the Euclidean time circle. The momentum assignments are opposite in the twisted sector ({\it i.e.} odd winding). This spectrum is mutually local and leads to a modular invariant torus amplitude.

The spectrum \thovect\ is invariant under $R \to {1 \over R}$ (accompanied by the exchange $2n \leftrightarrow w$) up to charge conjugation, {\it i.e.} the exchange of S and C conjugacy classes. Thus, this theory is self-dual at $R=1$.

The T-duality symmetry of this theory follows from enhanced gauge symmetry \DineVU. At the self-dual point $R=1$, an additional set of discrete state gauge bosons become physical,
\eqn\vectgauge{
A_\pm^I = e^{-\varphi} \psi_x \bar{\lambda}^I e^{\pm i \bar{x}}
}
The above vertex operators enhance the gauge symmetry from $Spin(24) \times U(1)$ to $Spin(26)$.

For the gauge symmetry enhancement to be consistent, the representations of $Spin(24)$ must extend as well. There are states which become massless at $R=1$ with vertex operators
\eqn\vectmassless{
\wt \Psi_0^{\pm} = e^{-\varphi /2 - i H /2} e^{\pm i \bar{x} } e^{\phi
+\bar{\phi}}
}
Here we mean ``massless'' in the sense of one-dimensional Liouville theory \SeibergEB; they have Liouville factor $e^{\phi+\bar{\phi}}$. This concept of masslessness coincides with loss of exponential damping in the partition function and so is quite natural. The $\wt \Psi_0^{\pm}$ combine with $\wt{\Psi}_V$ into the fundamental representation of $Spin(26)$. Likewise, $T_C$ and $T_S$ combine into a $Spin(26)$ spinor (as does $\Psi_C$ and $\Psi_S$).

As a final point of interest, note that there is a point of ``$(1,0)$ space supersymmetry''. At $R=1$, there is a gravitino ${\wt \Psi}_0 (n=w=-1) = {\wt S}^- \bar{\cal J}$ where
\eqn\sminus{
{\wt S}^- = e^{-\half\varphi - i\half H } V_{-\half, -1} = e^{-\half\varphi - i\half H -i x}
}
It is only at $R=1$ that ${\wt S}^-$ is a weight $(1,0)$ current. This is a special case of the construction in \KutasovUA.

The torus amplitude for this theory is
\eqn\vectpart{
Z_{T^2} = \cases{ -{V_L \over 2} \left(R + {2 \over R} \right) & $R>1$ \cr
         -{V_L \over 2} \left(2 R + {1 \over R} \right) & $R<1$
}}
The derivation of this result is discussed later in this paper. This function is continuous at $R=1$ but not analytic. The lack of smoothness is caused by the massless states \vectmassless\ and is indicative of a phase transition. Note that \vectpart\ respects the self-duality of this theory under $R \to { 1 \over R}$.

\subsec{$(-1)^{f_L}$ twist}

Twisting by $(-1)^{f_L}$ shifts the THO spectrum from \thocirc\ to
\eqn\thospin{\eqalign{
&T_V = e^{-\varphi} \bar{\cal O}_V(\half+n(2w+1)) V_{n,2w+1}
\cr
& T_C = e^{-\varphi} \bar{\cal O}_C (\half + (2n+1)w) V_{n+\half,2w} \cr
&\Psi_S = e^{-\varphi/2 + iH/2} \bar{\cal O}_S (\half + 2nw) V_{n,2w}~,
\quad\quad\quad\quad\quad\quad~~~~~~ p_R\geq 0~~\cr
&\Psi_0 =e^{-\varphi/2 + iH/2} \bar{\cal O}_0 (\half + (n+\half)(2w+1))
V_{n+{1\over 2},2w+1}~,~~~p_R\geq 0~~\cr
&\wt\Psi_C =e^{-\varphi/2 - iH/2} \bar{\cal O}_C (\half + n(2w+1))
V_{n,2w+1}~,\quad\quad\quad
~~~~ p_R\leq 0\cr
&\wt\Psi_V =e^{-\varphi/2 - iH/2} \bar{\cal O}_V (\half +
(2n+1)w)V_{n+\half,2w}~,\quad\quad\quad
~~~p_R\leq 0
}}
Similar to the $(-1)^{F_L}$ twisted theory, the even winding sector contains states with integer (half-integer) momenta that are even (odd) under $(-1)^{f_L}$ and oppositely for the twisted sector.

There are again 1-d Liouville massless states at $R=1$ with vertex operators
\eqn\spinmassless{
\Psi_0^{\pm} = e^{-\varphi /2 + i H /2} e^{\pm i \bar{x} } e^{\phi
+\bar{\phi}}
}
however there is no enhancement of the gauge symmetry to ensure T-duality. In fact, this theory has no T-duality symmetry.

Additionally, like the $(-1)^{F_L}$ twist, there is a point of ``supersymmetry''.  In this case, the gravitino is of opposite chirality; at $R=1$ there is an operator ${\Psi}_0 (n=w=0) = {S}^+ \bar{\cal J}$ where
\eqn\splus{
{S}^+ = e^{-\half\varphi + i\half H } V_{\half, 1} = e^{-\half\varphi + i\half H +i x}
}
Again, this is an example of the operators discussed in \KutasovUA.

The torus amplitude for this theory is
\eqn\spinpart{
Z_{T^2} =\cases{ -{V_L \over 2} \left(R - {1 \over R} \right) & $R>1$ \cr
         -V_L  \left( R - {1 \over R} \right) & $R<1$
}}
This computation is discussed more in a later section. Note this theory is T-dual under $R \to { 1 \over R}$ to the HO theory with $(-1)^{f_L}$ twisted compactification (see \DavisQE) and that \spinpart\ respects this duality.

\subsec{Thermal THO}

Now we come to the finite temperature version of the THO theory. This is obtained by adding a twist by the spacetime fermion number $(-1)^{F_R}$ as one traverses the Euclidean time circle. Recall from section 3.3 that $(-)^{F_R}=(-)^{F_L+f_L}$ by the GSO projection for the THO theory. So this theory can be thought of as twisted by $(-)^{F_L+f_L}$, the final element in the center of the gauge group. The spectrum is as follows:
\eqn\thotherm{\eqalign{
  &T_C=e^{-\varphi}\bar \CO_C(\half +2nw) V_{n,2w}
  \cr
  &T_0=e^{-\varphi}\bar \CO_0(\half +(n+\half)(2w+1)) V_{n+\half,2w+1}
  \cr
  &\Psi_S = e^{-\half\varphi + i\half H }\bar \CO_S(
  \half +(2n+1)w) V_{n+\half,2w}~,\qquad ~~p_R \geq 0  \cr
  &\Psi_V = e^{-\half\varphi + i\half H }\bar \CO_V(
  \half +n(2w+1)) V_{n,2w+1}~,\qquad ~~p_R \geq 0  \cr
  &\wt \Psi_S = e^{-\half\varphi - i\half H }\bar \CO_S(
  \half +n(2w+1)) V_{n,2w+1}~, \qquad ~~p_R \leq 0  \cr
  &\wt \Psi_V = e^{-\half\varphi - i\half H }
  \bar \CO_V(\half +(2n+1)w) V_{n+\half,2w}~, \qquad ~p_R \leq 0 \cr
  }}
Even winding operators have appropriate Matsubara frequencies: integer (half-integer) moding around the thermal circle for bosonic (fermionic) particles. The odd winding, or twisted, sector has opposite momentum assignments.

Note that the spectrum \thotherm\ is T-dual under $R\to {1 \over R}$ (with $2n \leftrightarrow w$), up to spacetime parity. However, this is not guaranteed by an enhancement of gauge symmetry at the self-dual point; that is, there are no extra physical gauge bosons at $R=1$. Although there is no enhanced symmetry, there are massless states at $R=1$ with vertex operators
\eqn\thermassless{
T_0^\pm = e^{-\varphi}V_{\pm\half,\mp 1} = e^{-\varphi} e^{\pm i{\bar
x}} e^{\phi + {\bar\phi}}
}

The torus amplitude for this theory is
\eqn\thermpart{
Z_{T^2} =\cases{ -{V_L \over 2} \left(R - {1 \over R} \right) & $R>1$ \cr
                  {V_L \over 2} \left( R - {1 \over R} \right) & $R<1$
}}
The calculation of this quantity is discussed in section 6. Note that \thermpart\ respects the self-duality of this theory under $R \to { 1 \over R}$. However, since this symmetry is not part of a gauge symmetry, it is not guaranteed at the non-perturbative level.

The thermodynamics implied by \thermpart\ are peculiar. Using the usual relation between Euclidean time radius and temperature, $R=(2\pi T)^{-1}$, we obtain
\eqn\free{
F =\cases{ {V_L } \left({1 \over 4 \pi} - \pi T^2 \right) & $T<(2\pi)^{-1}$ \cr
                 - {V_L } \left({1 \over 4 \pi} - \pi T^2 \right) & $T>(2\pi)^{-1}$
}}
This consists of a vacuum energy and a field theoretic piece which goes as $T^2$. For low temperatures \free\ is a sensible result and the field theoretic term is that of 24 free chiral fermions, as expected. On the other hand, the high temperature result is counter to conventional thermodynamic concepts; in particular, the expression \free\ leads to negative entropy for $T>(2\pi)^{-1}$ and a negative latent heat for the transition at $T=(2\pi)^{-1}$. Similar problems occur in the HO and HE theories \DavisQE. For more discussion on the interpretation of these thermal theories, see \SeibergNK.

\newsec{Moduli Space}

In \DavisQE\ it was found that the HO, HE and THO theories lie on a 13-dimensional moduli space parameterized by the radius of Euclidean time and a 12 component Wilson line (we will use $R$ and the radius-scaled gauge field $R\vec{A}$ as coordinates on moduli space). The Wilson line has the physical effect of shifting the momentum and lattice of gauge quantum numbers. Specifically, an operator with quantum numbers $(n,w,\vec{\omega})$ in the untwisted circle compactification will, in the presence of Wilson line $R\vec{A}$, have these numbers shifted to
\eqn\shift{
\eqalign{
n &\to n - \vec{\omega} \cdot R\vec{A} +{w\over 2} (R\vec{A})^2 \cr
w & \to w \cr
\vec{\omega} &\to \vec{\omega} - wR\vec{A}
  }}
This shift can be understood as a $SO(1,13)$ transformation on the lattice of allowed $(n,w,\vec{\omega})$. This does not affect mutual locality or modular invariance, {\it i.e.} the lattice remains even and self-dual under this transformation. As such it does not interfere with the physical conditions from BRST quantization. Thus, if an operator is physical with the quantum numbers $(n,w,\vec{\omega})$, it will be physical with the shifted quantum numbers \shift.

There is an $SO(13)$ subgroup of the above transformations which leaves the spectrum of masses invariant. Thus the moduli space is locally $SO(1,13) / SO(13)$. This is only true locally since a number of discrete identifications must be made on moduli space ({\it e.g.} through T-duality or adding $R\vec{A} \in 0$). Arguments from the critical string suggest that the exact moduli space is $SO(1,13, \Bbb Z) \backslash SO(1,13) / SO(13)$, but the situation here is not clear. One distinction between the $d=2$ heterotic strings and the critical heterotic strings which may be of importance here is that there are only two distinct decompactification limits in the $d=10$ case, {\it i.e.} $Spin(32)/ \Bbb Z_2$ and $E_8 \times E_8$, while there are three such limits in two-dimensions (the HE, HO and THO theories). We will conservatively refer to the moduli space as ${\cal M} = {\cal H} \backslash SO(1,13) / SO(13)$ where ${\cal H}$ is an unknown discrete group.

Before moving on to the phase structure of moduli space let us single out the most symmetric lines of theories. In \DavisQE\ it was observed that the compactifications of HO and HE with twists by the center of the gauge group could be understood as compactifications with Wilson lines. The same is true of the THO theory. The $(-1)^{F_L}$ twist can be implemented by adding $R\vec{A} =(1,0^{11}) \in V$, the $(-1)^{f_L}$ by $R\vec{A}=(\half^{12}) \in S$, and a Wilson line $R\vec{A}=(-\half,\half^{11}) \in C$ implements the $(-1)^{F_L+f_L}$ twist. Hence, all of the compactifications discussed in the previous section are continuously connected.

\subsec{Massless Particles and Phase Transitions}
As previously discussed, the appearance of massless particles at some point in moduli space is indicative of a phase transition at this point. Thus it is of interest to map out the subspaces which support massless quanta. To this end, consider the vertex operator
\eqn\totvert{
{\cal V} = \CO_{g.s.} e^{i\vec{\omega}\cdot\vec{H}_L}
e^{i p_R x + ip_L {\bar x}} e^{(1-|p_R|) (\phi +{\bar\phi})}
}
where $p_{R,L} = {n\over R} \pm {wR\over 2}$ and there is no Wilson line turned on. The notation $\CO_{g.s.}$ refers to the ground state operator (with $\Delta=\half$) of the right-moving side in canonical ghost picture, {\it i.e.} $e^{-\varphi}$ or $e^{-{\varphi \over 2} \pm i {H \over 2}}$. The right-moving factor of the wave function contributes $\Delta_R={1\over 2}$ for all $p_R$. The physical condition ($\bar{L}_0 =1$) for the left-moving side is then
\eqn\physcond{
\vec{\omega}^2 = 1 + 2nw
}
Starting from a physical state we turn on a Wilson line. This shifts the momenta while preserving the physical condition.

The massless states in the theory with Wilson line are those satisfying $\wt{p}_R={n-\vec{\omega} \cdot R\vec{A} + {w \over 2}(R\vec{A})^2 \over R} + {wR \over 2}=0$. Using the physical condition \physcond, this can be expressed as
\eqn\mcirc{
\left( R\vec{A} - {1\over w} \vec{\omega} \right)^2 + R^2 = {1\over w^2}
}
Since $R \ge 0$, this is a hemisphere in moduli space with center at $(R,R\vec{A}) = (0,{1\over w} \vec{\omega})$ and radius $1/|w|$. The conclusion is valid independently of which state $(n,w,\vec{\omega})$
we started with, as long as $w\neq 0$. For states with zero winding the massless
states correspond instead to the half-plane
\eqn\wzerocon{
\vec{\omega} \cdot R\vec{A} = n
}
in moduli space. Note that this discussion is valid for any of the three theories. The difference between the theories lies in which $\CO_{g.s.}$ is paired with a given $(n,w,\vec{\omega})$ and what is the most natural way to parameterize moduli space.

Let us illustrate some aspects of the phase structure with a plot. In Fig. 1 we see a cross-section (the $R A_{k \ne 1}=0$ plane) of some of the hemispheres supporting massless particles. We have chosen coordinates on moduli space such that $(R=\infty, R\vec{A}=0)$ is the non-compact HO theory. Some key features:

\item{i.} The line $R\vec{A}=(1,0^{11})$ is the thermal HO theory. Thus the topmost curve represents the thermal phase transition occurring at $R=1$ discussed in \DavisQE. The quantum numbers of the particles which become massless along this curve are $(n=0, w=\pm 1, \vec{\omega}=(\pm 1, 0^{11}))$. Note that the radius where this transition occurs varies such that at $R A_1=0$ there is no transition at finite $R$. In fact, none of the curves plotted cross $R\vec{A}=0$ at finite $R$. This is consistent with the analysis of \DavisQE\ where the circle compactified theory was found to have no transition.

\item{ii.} Note the intricacy of the structure: we see that the distribution of these surfaces becomes very dense at small $R$ except along the vertical line $R A_1 =1$. When we compute the torus amplitude in the following section, this will prevent us from finding a general expression at small radius, {\it except} for special Wilson lines such as $R A_1 =1$.

\item{iii.} There are examples of intersecting surfaces. Physically, this means that several sets of particles become massless at the same point.

\item{iv.} These surfaces all correspond to \mcirc\ with some $\vec{\omega} \in V$. Thus, since we are discussing the HO theory, the particles becoming massless are spacetime scalars. If we had chosen coordinates such that $(R=\infty, R\vec{A}=0)$ is the THO theory, we would generate an identical plot but the massless particles would be spacetime fermions. There are also families of surfaces with $\vec{\omega}$ in the other weight lattices.

\medskip \ifig\figI{Plot of some phase transition surfaces in ${\cal M}$. Horizontal axis is one component of the Wilson line and all others have been set to zero, {\it i.e.} $R\vec{A}=(RA,0^{11})$.} {\epsfxsize=0.6\hsize\epsfbox{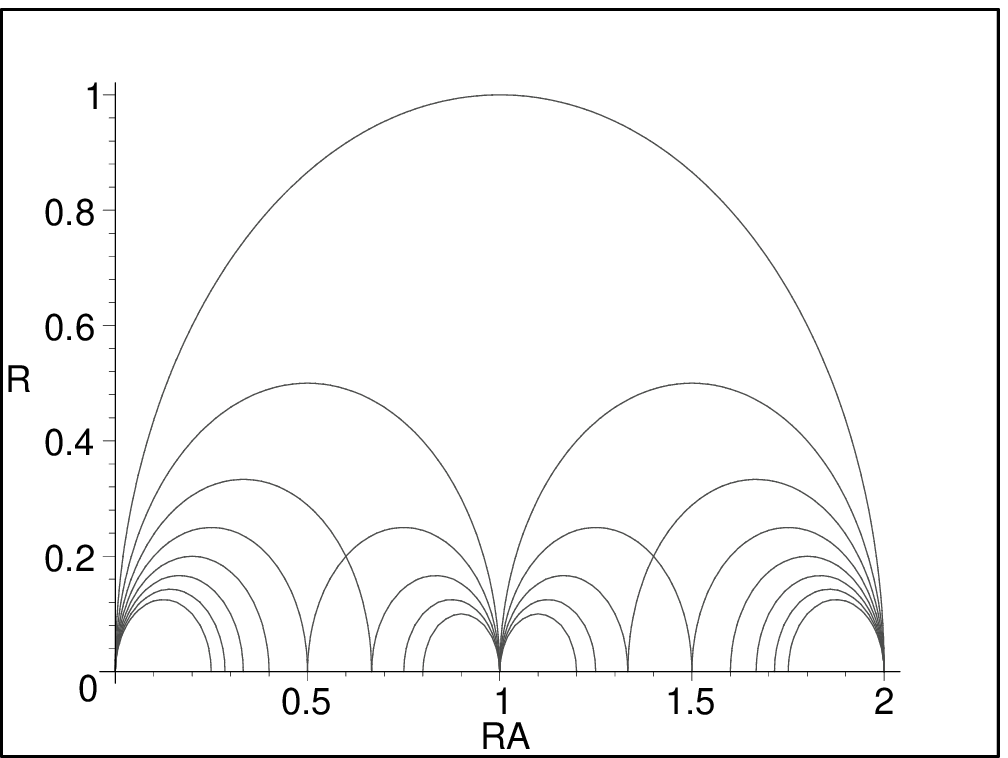}}

\subsec{Comment on Discrete States}

As is well-known \refs{\ItohIY\BouwknegtVA-\BouwknegtAM}, the BRST cohomology of $d=2$ NSR string is not entirely trivial at higher level; there are the so-called ``discrete states''. The results of this BRST analysis can be applied in a straightforward manner to the heterotic string. To summarize, in addition to the propagating spectrum there exist operators parameterized by two positive integers $(r,s)$:
\eqn\discrete{
{\cal V}_{r,s} = \CO_{g.s.} \CO_{r,s} e^{i\vec{\omega}\cdot\vec{H}_L}
e^{i \left( {r-s \over 2} \right) x + ip_L {\bar x}} e^{(1-{r+s \over 2} ) (\phi +{\bar\phi})}
}
where $\CO_{r,s}$ is a certain level ${rs \over 2}$ operator composed of oscillator excitations of the right-moving CFT and $p_L = {r-s \over 2} -wR$. The $\CO_{g.s.}$ is one of the canonical ground state operators discussed in the previous subsection, but note that ${\cal V}_{r,s}$ is in the NS (R) sector if $r-s$ is even (odd). The above operator is only physical on select subspaces in moduli space where ${r-s \over 2}$ is a physical momentum. Using the condition $\bar{L}_0=1$, this hypersurface can be shown to be
\eqn\discmod{
\left(R \vec{A} - {1\over w} \vec{\omega} \right)^2 + \left[ R - \left({r-s} \over 2 w \right) \right]^2 = { 1 + \left({r+s \over 2}\right)^2 \over w^2}
}
for operators with non-zero winding. Operators with $w=0$ are physical on the surface
\eqn\discwzero{
{r-s \over 2} = { n - \vec{\omega} \cdot R\vec{A} \over R}
}
All of the enhanced gauge bosons and other discrete states discussed in this paper as well as in \refs{\DavisQE, \SeibergNK} are examples of the operators here. It would be of interest to further investigate the spectrum of discrete states of $d=2$ heterotic string theory.

\newsec{The Torus Partition Function}

We wish to calculate the extensive portion of the partition function, that is, the contribution from the region of weak coupling. Consider first a theory with compactified Euclidean time (no twist). The extensive piece of the partition function is given by the 1-loop torus amplitude,
\eqn\torus{
Z_{T^2} (R) = \int_{\cal F} { d\tau d\bar{\tau} \over 4 \tau_2} {V_L \over \sqrt{8 \pi^2 \tau_2} } \eta({\bar\tau})^{-12} \sum_{\vec{\omega}} \sum_{n,w \in \bz} \bar{q}^{\half \vec{\omega}^2} q^{\half \left({n \over R} + {wR \over 2}\right)^2} \bar{q}^{\half \left({n \over R} - {wR \over 2}\right)^2}
}
where the nome $q$ is defined $q= \exp( 2 \pi i \tau)$. The factor $V_L (8 \pi^2 \tau_2 )^{-\half}$ comes from the integration over the Liouville zero-mode. The contribution from the oscillators of the embedding coordinates and their superpartners are cancelled by those of the ghost and superconformal ghost systems on the worldsheet. The current algebra contributes the sum over lattice vectors $\vec{\omega}$ and the eta-function factor. The range of integration is the fundamental modular domain, $\CF = \{ |\tau | >1 , |\tau_1| < \half, \tau_2 >0 \}$.

In the above, we have not specified the range of the sum over lattice vectors $\vec{\omega}$. For the HO theory, the $\vec{\omega}$ sum\foot{The coefficients of $-\half$'s for a few of these sums are due to fermionic statistics and the constraint that fermions have ${\rm sign}(p_R) = (-)^{f_L}$.} is
\eqn\hosum{
\left[ \sum_{\vec{\omega}\in V} - \half\sum_{\vec{\omega}\in S\oplus C}\right]
}
while for the THO theory the sum is
\eqn\thosum{
 - \half \left[\sum_{\vec{\omega} \in C} -\half \sum_{\vec{\omega} \in S \oplus V}  \right]
}
In the above expressions $V, S, C$ are $Spin(24)$ weight lattices. For the $Spin(8)\times E_8$ HE theory, split the lattice vector into two vectors of 4 and 8 components: $\vec{\omega} = (\vec{\omega}^{(1)}, \vec{\omega}^{(2)})$. The relevant sum is then
\eqn\hesum{
\left[ \sum_{\vec{\omega}^{(1)}\in V} - \half \sum_{\vec{\omega}^{(1)}\in S\oplus C}\right] \sum_{\vec{\omega}^{(2)} \in 0 \oplus S}
}
Here $0,V,S,C$ denote $Spin(8)$ and $Spin(16) \subseteq E_8$ weight lattices for the $\vec{\omega}^{(1)}$ and $\vec{\omega}^{(2)}$ sums, respectively. Note that these sums are such that they reproduce the results of section 3. For example,
\eqn\examplesum{
\left[ \sum_{\vec{\omega}\in V} - \half \sum_{\vec{\omega}\in S\oplus C}\right] \bar{q}^{\half \vec{\omega}^2} = Z_{HO}^{(\lambda)} (\bar{\tau})
}
where $Z_{HO}^{(\lambda)}$ is defined in \hopart.

\subsec{Wilson lines}

The partition function with Wilson lines is obtained by shifting the sums in \torus\ as indicated by \shift. This results in
\eqn\partfct{
Z_{T^2} (R, R\vec{A})= \int_{\cal F} {d\tau d{\bar\tau}\over 4\tau_2} ~ {V_L\over\sqrt{8\pi^2\tau_2}}
\eta({\bar\tau})^{-12} \sum_{\vec{\omega}} \sum_{n,w\in \bz}
{\bar q}^{{1\over 2}\left(\vec{\omega} - w R \vec{A}\right)^2}
{\bar q}^{{1\over 2}{\tilde{p}}^2_L} q^{{1\over 2}{\tilde{p}}^2_R}
}
where $\tilde{p}_{R,L}= {n-\vec{\omega}\cdot R \vec{A} + \half w (R \vec{A})^2 \over R} \pm {w R \over 2}$ with $n,w \in \bz$.

We will proceed to perform the modular integral in \partfct\ and obtain a rather simple result, at least for sufficiently large $R$. One can perform Poisson resummation over $n$ in \partfct\ using the identity
\eqn\rsum{ \sum_{n\in \bz +\nu} e^{-\pi a n^2 + 2\pi i b n}
= {1\over\sqrt{a}} \sum_{m\in \bz} e^{2\pi i \nu m} e^{-\pi (m-b)^2/a}
}
to obtain
\eqn\resumn{
Z_{T^2}= {R V_L \over 16 \pi} \int_{\cal F} {d\tau d{\bar\tau}\over \tau_2^2 \eta(\bar{\tau})^{12}} \sum_{m,w\in \bz} e^{-S(m,w)} e^{- i \pi mw (R \vec{A})^2} \sum_{\vec{\omega}} \bar{q}^{\half ( \vec{\omega} - w R \vec{A})^2} e^{-2 \pi i m R \vec{A} \cdot (\vec{\omega} - w R \vec{A} )}
}
The worldsheet instanton factor is $S(m,w) = {\pi R^2\over 2\tau_2} | m - w\tau |^2$. At this point, subtleties regarding the convergence of this expression arise. These subtleties will be addressed in section 6.3. For now, we press on under the assumption that \resumn\ is absolutely convergent.

Now that the sum over $\vec{\omega}$ has been sufficiently disentangled from the other sums, we go ahead and evaluate it. The results for our three theories are:
\item{i.}
{\bf HO}:
\eqn\hoint{\eqalign{
Z_{HO}(R,R\vec{A}) & =
{RV_L\over 32\pi}\int_{\cal F}
{d\tau d\bar{\tau}\over \tau_2^2}
\sum_{m,w\in \bz} e^{-S(m,w)}
e^{-i\pi mw (R\vec{A})^2}
\left[
\prod_{i=1}^{12} Z_{2mRA_i}^{-2wRA_i}(\tau)^* \right. \cr
&~ \left. - \prod_{i=1}^{12} e^{-i\pi wR A_i} Z_{1+2mRA_i}^{-2wRA_i}(\tau)^*
 -  \prod_{i=1}^{12} Z_{2mRA_i}^{1-2wRA_i}(\tau)^*\right]
 }}

\item{ii.}
{\bf HE}:
\eqn\heint{\eqalign{
Z_{HE}(R,R\vec{A}) & =
{RV_L\over 64\pi}\int_{\cal F}
{d\tau d\bar{\tau}\over \tau_2^2}
\sum_{m,w\in \bz} e^{-S(m,w)}
e^{-i\pi mw (R\vec{A})^2 }
\left[
\prod_{i=1}^{4} Z_{2mRA_i}^{-2wRA_i}(\tau)^* \right. \cr
&~ \left. - \prod_{i=1}^{4} e^{-i\pi wR A_i} Z_{1+2mRA_i}^{-2wRA_i}(\tau)^*
 - \prod_{i=1}^{4} Z_{2mRA_i}^{1-2wRA_i}(\tau)^*\right] \cr
&~ \times \left[ \prod_{i=5}^{12} Z_{2mRA_i}^{-2wRA_i}(\tau)^*
+ \prod_{i=5}^{12} e^{-i\pi w R A_i} Z_{1+2mRA_i}^{-2wRA_i}(\tau)^* \right. \cr
&~ \left. \quad\quad + \prod_{i=5}^{12} Z_{2mRA_i}^{1-2wRA_i}(\tau)^*
 +  \prod_{i=5}^{12} e^{-i\pi(w R A_i -1/2)} Z_{1+2mRA_i}^{1-2wRA_i}(\tau)^* \right]
 }}

\item{iii.}
{\bf THO}:
\eqn\thoint{\eqalign{
Z_{THO}(R,R\vec{A})  &= -
{RV_L\over 64\pi} \int_{\cal F}
{d\tau d\bar{\tau}\over \tau_2^2}
\sum_{m,w\in \bz} e^{-S(m,w)}
e^{-i\pi mw (R\vec{A})^2} \cr
&~~~~\times \left[ \prod_{i=1}^{12} Z^{-2wRA_i}_{2mRA_i} (\tau)^* - \prod_{i=1}^{12} e^{-i\pi wRA_i} Z^{-2wRA_i}_{1+2mRA_i} (\tau)^* \right. \cr
 &~~~~~~~~- \left. \prod_{i=1}^{12} Z^{1-2wRA_i}_{2mRA_i} (\tau)^* + 3 \prod_{i=1}^{12} e^{-i \pi w RA_i} Z^{1-2wRA_i}_{1+2mRA_i} (\tau)^* \right]
}}

The notation $Z^\alpha_{~\beta}$ was introduced in \PolchinskiRR. These expressions can be shown to be modular invariant\foot{The modular properties are
(note that these correct formula 10.7.14 in \PolchinskiRR)
\eqn\zprop{
\eqalign{
Z^{\alpha}_\beta (\tau + 1) & =
e^{-i\pi [ \alpha/2 (\alpha/2+1) + 1/12]}  Z^\alpha_{\alpha+ \beta+ 1}  (\tau) \cr
Z^{\alpha}_\beta (-1/\tau) & = e^{i\pi\alpha\beta/2} Z^\beta_{-\alpha} (\tau)
}}}.

To perform the $\tau$-integral we use the trick established in \refs{\PolchinskiZF,\AlvarezSJ,\BrienPN}. The integral over the fundamental domain may be written as an integral over the strip, $E=\{ |\tau_1 | < \half, \tau_2>0\}$, as follows:
\eqn\ftoe{
\int_{\CF} \sum_{m, w \in \bz}  \Gamma (m,w) e^{-S(m,w)} = \int_{\CF} \Gamma(0,0) + \int_E \sum_{m \ne 0}  \Gamma(m,0) \, e^{-S(m,0)}
}
The above identity holds for any $\Gamma(m,w)$ such that the left-hand side is modular invariant. Performing this explicitly for the HO theory,
\eqn\zpoisII{\eqalign{
& Z_{HO}(R,R\vec{A})  =
{RV_L\over 32\pi}\left\{ \int_{\cal F}
{d\tau d\bar{\tau}\over \tau_2^2}
\left[
\prod_{i=1}^{12} Z_{0}^{0}(\tau) - \prod_{i=1}^{12} Z_{1}^{0}(\tau)
 -  \prod_{i=1}^{12} Z_{0}^{1}(\tau)\right]^* \right. \cr
&+ \left. \int_E
{d\tau d\bar{\tau}\over \tau_2^2}
\sum_{m \ne 0} e^{-S(m,0)}
\left[
\prod_{i=1}^{12} Z_{2mRA_i}^{0}(\tau) - \prod_{i=1}^{12}  Z_{1+2mRA_i}^{0}(\tau)
 -  \prod_{i=1}^{12} Z_{2mRA_i}^{1}(\tau)\right]^* \right\}
  }}
The first integral is simple to evaluate since $Z_{HO}^{(\lambda)} = 24$. The integral on the strip $E$ implements level matching. This means that only terms of order $q^0 \bar{q}^0$ survive the integration over $E$. Performing the modular integral with only these terms is straightforward and results in
\eqn\hopart{
Z_{HO}(R,R\vec{A})  = V_L \left\{ R + {1\over \pi^2 R} \sum_{i=1}^{12}
\sum_{m=1}^\infty {\cos(2\pi mRA_i) \over m^2} \right\}
}
One can use the Fourier expansion \eqn\fourt{ {1\over \pi^2}\sum_{m=1}^\infty
{\cos(2\pi mx) \over m^2} = |x|^2 - |x| + {1\over 6}, \quad\quad |x| \leq 1 } to simplify the result further,
\eqn\hofinal{
Z_{HO} (R,R\vec{A}) = V_L \left\{ R + {2 + (R \vec{A})^2 - \sum_{i=1}^{12} |R A_i| \over R} \right\}
}
for $|R A_i| \le 1$. Note that the result, \hopart, is
manifestly periodic under $RA_i\to RA_i + 1$ for each $i$. This periodicity and the result \hofinal\ are sufficient to evaluate $Z_{T^2}$ for all $R\vec{A}$.

For the HE theory we find
\eqn\hefinal{
Z_{HE}(R,R\vec{A})  = {V_L \over R} \left\{ -\sum_{i=1}^4 |R A_i| + {1 \over 8} \sum_{\left\{ \epsilon_k \right\} = \pm 1 } \left| \sum_{n=1}^4 \epsilon_n R A_n \right| \right\}
}
for $|R A_i| \le 1$; beyond that one can use the periodicity $Z_{HE}(R,RA_i+2) =  Z_{HE}(R,R A_i)$. The lack of dependence on the $E_8$ Wilson line is expected since there are no $E_8$ charged particles in the massless sector.

We can follow the same path with the THO theory to obtain
\eqn\thofinal{
Z_{THO}(R,R \vec{A}) = - {V_L \over 2} \left\{ R + {2 + (R\vec{A})^2 -\sum_{i=1}^{12}|R A_i| \over R} \right\}
}
This is valid for Wilson lines that satisfy $|R A_i| \le 1$ (with periodicity $Z_{THO}(R,RA_i+1) = Z_{THO}(R,R A_i)$).

\subsec{Field Theory Calculation}

In the above torus amplitudes we consistently find a term proportional to ${1 \over R}$ which is the contribution from field-theoretic modes. This term provides an interesting check on our results for it can be calculated purely from the low energy field theory as follows. The massless modes contribute a vacuum energy of $-{(-1)^{F_R} V_L |{\wt n}| \over R}$ where ${\wt n}$ is the shifted momentum quantum number (see \shift). Hence, we sum over all massless modes in the theory, {\it .i.e} all modes of zero winding,
\eqn\ft{
Z_{f.t.} = -{ V_L \over  R} \sum_{\vec{\omega}} \sum_{n \in \bz} (-1)^{F_R} \left| n - \vec{\omega} \cdot R \vec{A} \right|
}
The $\vec{\omega}$ sum depends on the theory and is given by \hosum, \thosum, or \hesum\ with the additional constraint $\vec{\omega}^2 =1$ (which is required from the physical condition for $w=0$). %The normalization in \ft\ is chosen so that for the thermal theory $Z_{f.t.} = -\beta F$, where $F$ is the free energy of a one-dimensional gas.

The sum over $n$ can be performed via $\zeta$-function regularization to obtain the ${1 \over R}$ terms above. For example, in the HO theory the sum becomes,
\eqn\hoft{\eqalign{
Z &= -{ 2 V_L \over  R} \sum_{i=1}^{12} \sum_{n=1}^{\infty} \left( n-|R A_i| \right) \cr
  &= { V_L \over  R} \left[ 2 + (R\vec{A})^2 -\sum_{i=1}^{12}|R A_i| \right]
}}
as in \hofinal. Likewise, one can also reproduce \hefinal\ and the ${1 \over R}$ term in \thofinal\ from the formula \ft.

\subsec{Small radius}

In the analysis of section 6.1, some subtleties were glossed over. While \partfct\ is absolutely convergent for all $R$ and $R\vec{A}$, the resummed expression \resumn\ can lose absolute convergence at small enough $R$. At large $\tau_2$, a given term in the integrand is exponentially damped provided that
\eqn\damp{
\left(R \vec{A} - {1 \over w } \vec{\omega}  \right)^2 + { R^2 \over 2}   > { 1 \over w^2}
}
Clearly, $R > \sqrt{2}$ is large enough to provide the needed damping for all values of winding, lattice vector and Wilson line (the exception $w=0$ does not cause a problem since the lattice sum is only over vectors with $\vec{\omega}^2 \ge 1$). At smaller values of the radius the integrals of individual terms in \resumn\ may diverge and so one cannot exchange the summation and integration order.

The physical meaning of this loss of convergence becomes clear when one compares the above expression with that defining the surface of massless particles. At some fixed $R\vec{A}$, define $R_{conv}$ as the largest radius such that \damp\ is violated for some particle in the spectrum and define $R_{trans}$ as the largest radius for which \mcirc\ is satisfied by some (not necessarily the same) mode. Then we see that at fixed Wilson line as one lowers $R$ from infinity there is loss of convergence at a radius slightly above (by a universal factor) the radius of phase transition, {\it i.e.} $R_{conv} = \sqrt{2} R_{trans}$. One may continue the results of the previous section all the way down to $R_{trans}$, where there is a non-analyticity due to the presence of massless modes. This occurs because the contribution of this mode is proportional to $|{\wt p}_R|$, which vanishes at the transition point.

In \DavisQE\ the torus amplitudes for HO and HE were calculated both above and below the transition for specific Wilson lines. For $R>R_{trans}$, the calculations were done just as in the previous section. Below the transition, one Poisson resums winding instead of momentum; this avoids the convergence issues just addressed. Note that this only works for special Wilson lines (those which do not break the gauge group). If one were to resum winding in \partfct\ with arbitrary $R\vec{A}$ one obtains an expression like \resumn\ but with very bad convergence properties; at generic $R\vec{A}$ the expression is not absolutely convergent for any $R$. This is caused by the dense distribution of phase transitions as illustrated by Fig. 1. As also seen in Fig. 1, there is only a single phase transition for the theory with $R \vec{A} = (1,0^{11})$. Thus the expression obtained after resummation of winding has a finite region of convergence.

\subsec{Discussion}

We see from the considerations of the last subsection, as well as those of section 5, that the moduli space of two-dimensional theories is quite rich. The theories at large $R$ are well understood and tractable. They have well-defined field theory limits and the torus amplitudes are computable in a straightforward fashion. However, a dense and intricate web of nested and intersecting surfaces of phase transitions exists at small $R$. This prevents one from computing the partition function as well as often obscuring the existence of a field theory limit as $R \to 0$. The situation might be clearer in another set of coordinates, perhaps related to $R$ and $R\vec{A}$ via a generalized T-duality\foot{I thank N. Seiberg for this suggestion.}. Such coordinates could perhaps be constructed  with knowledge of the discrete symmetry group $\CH$ discussed in section 5. It would be of interest to develop this further.

%\vskip 3.0cm
\bigskip
\noindent {\bf Acknowledgements:} \medskip \noindent

I would like to especially thank F. Larsen and N. Seiberg for early collaboration on this project as well as many useful discussions. I would also like to thank B. Burrington and P. de Medeiros for helpful discussions and comments on an advanced draft of this paper. This work was supported by a Rackham Predoctoral Fellowship at the University of Michigan.

\listrefs

\end